\documentclass[11pt]{article}

\begin{document}
\setlength{\textwidth}{130mm}\setlength{\textheight}{194mm}

\title{Dynamics and symmetries on the noncommutative plane\footnote{extended version of a lecture presented at the XXII. Max Born Symposium, Wroclaw, 27. -- 29. Sept. 2006}}

\author{Peter C. Stichel\thanks{
An der Krebskuhle 21,
33619 Bielefeld
\tt{e-mail: peter@physik.uni-bielefeld.de}}}
\date{\it Dedicated to the 70th birthday of Jerzy (Jurek) Lukierski}
\maketitle


\section{Introduction}

Sometime in 1996 Jurek visited us in Bielefeld and asked about a dynamical particle model 
showing the presence of the second central charge of the planar Galilean group. This was a new and 
very interesting question. Soon we agreed that such a model must necessarily contain higher order time 
derivatives of the particle coordinates within a Lagrangian. After Bielefeld Jurek visited Wojtek Zakrzewski
 in Durham, and we had a long and intense discussion over several weeks by 
exchanging a lot of e-mails and faxes. In this way our first common paper [1] on
 the twofold centrally extended Galilean group and noncommutative geometry was born. But besides that
 it was the beginning of a deep friendship and successful collaboration that has continued up to now.

The aim of the present paper is to highlight the main results of our
common work on nonrelativistic particle models on the noncommutative plane and round them 
off by some new results. Therefore this paper is neither a review of the whole field, nor a critical comparison with the work done by other colleagues (I apologize to all whose papers will not be cited in the following).

The following considerations will concentrate on the noncommutative aspects of classical mechanics. Neither quantum effects nor quantum field theory will be considered.

\section{Lagrangian for a free exotic particle}

A particle characterized by the two central charges $m$ and $-\Theta$ of the planar
 Galilean group will be called exotic. $m$ and $\Theta$ appear in the following Lie-brackets
 (represented by Poisson-brackets (PBs)) between the translation generators $P_i$ and the boost generators $K_i$
\begin{equation}
\{ P_i, K_j \} = m\delta_{ij}\ , \quad \{ K_i, K_j\} = - \Theta \epsilon_{ij}\ .
\end{equation}
In order to find a Lagrangian whose Noether charges for boosts satisfy (1) we must add the second time derivative of the coordinates $\ddot{x}_i$ to the usual variables $x_i$ and $\dot{x}_i$. As shown in [1] the most general one-particle Lagrangian, which is at most linearly dependent on $\ddot{x}_i$, leading to the Euler-Lagrange equations of motion which are covariant w.r.t. the planar Galilei group, is given, up to gauge transformations, by
\begin{equation}
{\cal L} = \frac{m}{2} \dot{x}_i^2 + \frac{\Theta}{2} \epsilon_{ij} \dot{x}_i \ddot{x}_i\ .
\end{equation}
Using the 1st-order formalism (2) may be rewritten as
\begin{equation}
{\cal L} = P_i \dot{x}_i + \frac{\Theta}{2} \epsilon_{ij} y_i \dot{y}_i - H ({\bf y}, {\bf P})
\end{equation}
with
\begin{equation}
H({\bf y}, {\bf P}) = y_i P_i - \frac{m}{2} y_i^2\ .
\end{equation}
(3) describes a constrained system, because we have
\begin{equation}
\frac{\partial {\cal L}}{\partial \dot{y}_i} = - \frac{\Theta}{2} \epsilon_{ij} y_j\ .
\end{equation}
Therefore the PBs, obtained by means of the Faddeev-Jackiw procedure, take a non-standard form
\begin{equation}
\{ x_i, P_j\} = \delta_{ij}\ , \quad \{ y_i, y_j\} = - \frac{1}{\Theta} \epsilon_{ij}\ .
\end{equation}
All other PBs vanish.

For the conserved boost generator we obtain
\begin{equation}
K_i = - m x_i - \Theta \epsilon_{ij} y_j + P_i t
\end{equation}
and therefore, due to (6), the PB resp.~commutator of two boosts is nonvanishing
\begin{equation}
\{ K_i, K_j \} = - \Theta \epsilon_{ij}\ .
\end{equation}
The Lagrangian (3) shows that the phase space is 6-dimensional. In order to split off
 two internal degrees of freedom, we have to look for a Galilean invariant decomposition of the 6-dim phase space into two dynamically independent parts: a 4-dim external and a 2-dim internal part. This decomposition is achieved by the transformation ([1], [2]) $ ({\bf x}, {\bf P, y}) \to ({\bf X, P, Q})$
with
$$
y_i = \frac{P_i}{m} + \frac{Q_i}{\Theta}
$$
and
\begin{equation}
x_i = X_i - \epsilon_{ij} \frac{Q_j}{m}
\end{equation}
leading to the following decomposition of the Lagrangian (3)
$$
{\cal L} = {\cal L}_{ext} + {\cal L}_{int}
$$
with
$$
{\cal L}_{ext} = P_i \dot{X}_i + \frac{\Theta}{2m^2} \epsilon_{ij} P_i \dot{P}_j - \frac{P_i^2}{2m}
$$
\begin{equation}
{\cal L}_{int} = \frac{m}{2\Theta^2} Q^2_i + \frac{1}{2\Theta} \epsilon_{ij} Q_i \dot{Q}_j\ .
\end{equation}
From (9) and the PBs (6) it now follows that the new coordinates $X_i$ are noncommutative
\begin{equation}
\{ X_i, X_j \} = \frac{\Theta}{m^2} \epsilon_{ij}\ .
\end{equation}
The remaining nonvanishing PBs are
\begin{equation}
\{ X_i, P_j \} = \delta_{ij}\ , \quad \{Q_i, Q_j \} = - \Theta \epsilon_{ij}\ .
\end{equation}

\bigskip
\noindent
\underline{Conclusion:} The particle Lagrangian (2) containing $\ddot{x}_i$ leads to 
a nonvanishing commutator of two boosts. But in order to obtain noncommutative coordinates we are forced to decompose the 6-dim phase space in a Galilean invariant manner into two dynamically independent 4-dim external and 2-dim internal phase spaces. 

\bigskip
\noindent
\underline{Generalization of (2):} If we add to (2) a term $f (\ddot{x}_i^2)$, the 
obtained Lagrangian is the most general one involving, in a Galilean quasi-invariant manner, the variables $x_i, \dot{x}_i$ and $\ddot{x}_i$. 

Then one can show\footnote{Details will be published elsewhere}
\begin{description}
\item{i)} the PB of the two boosts (8) will not change,
\item{ii)} the new 8-dim phase space may be decomposed again in a Galilean invariant manner into two dynamically independent 4-dim parts, an external and an internal one.
\end{description} 

\section{Commutative - versus noncommutative plane}

The subalgebra of the Galilean algebra containing only translations and boosts is given 
in the cases of, respectively, their one- or two-fold central extensions by

\medskip
\hspace{0,5cm} \underline{one-fold centrally extended} \hspace{1,5cm} \underline{two-fold centrally extended}

\[
\begin{array}{ll}
\{ P_i, K_j \} = m \delta_{ij} \qquad  \qquad \qquad \qquad &\{ P_i^\prime, K_j^\prime \} = m \delta_{ij}\\
\{ P_i, P_j\}\ = 0  & \{ P_i^\prime,  P^\prime_j \} \ = 0\\
\{ K_i, K_j \} = 0 & \{ K_i^\prime,  K_j^\prime \}  = - \Theta \epsilon_{ij}
\end{array}
\]

Obviously both are related by the transformations
\begin{equation}
K^\prime_i = K_i - \frac{\Theta}{2m} \epsilon_{ij} P_j, \ \ \ P_i^\prime = P_i\ .
\end{equation}
To this corresponds the following point transformation between noncommutative coordinates $X_i$ and commutative ones $q_i$
\begin{equation}
X_i = q_i - \frac{\Theta}{2m^2} \epsilon_{ij} P_j
\end{equation}
as can be read off immediately from the form of ${\cal L}_{ext}$ in (10).

\noindent
Now the question arises: {\it What to use in physics, the commutative or the non-commutative plane?}

\noindent
\underline{Answer:} For free particles both possibilities are equivalent.
 But in the case of a nontrivial interaction one has to use the commutative 
(noncommutative) plane, if a local potential or gauge interaction is given in terms of $q_i$ $(X_i)$.

\section{General form of noncommutative mechanics}

Up to now noncommutativity has been described by a constant $\Theta$ in the PB (11). But it
 is possible to get $\Theta$ as a function of ${\bf X}$ and ${\bf P}$ if one considers
external Lagrangians more general than (10). 

To do this consider a very general class of Lagrangians given by 
\begin{equation}
{\cal L} = P_i \dot{X}_i + \tilde{A}_i ({\bf X}, {\bf P}) \dot{P}_i - H({\bf P}, {\bf X})
\end{equation}
leading to the PBs
\begin{equation}
\{ X_i, X_j \} \sim \epsilon_{ij} \tilde{B}, \ \ \tilde{B} : = \epsilon_{k\ell} \partial_{P_k} \tilde{A}_\ell ({\bf X}, {\bf P})
\end{equation}
with 
$$
\{ P_i, P_j\} = 0\ .
$$
We dispense with the reproduction of the more complicated form of $\{ X_i, P_j\}$.
 Again by the point transformation 
\begin{equation}
X_i \to q_i = X_i - \tilde{A}_i ({\bf X}, {\bf P})
\end{equation}
we obtain commuting coordinates $q_i$ as follows from
$$
P_i \dot{X}_i + \tilde{A}_i \dot{P}_i = P_i \dot{q}_i + \frac{d}{dt} (\tilde{A}_i P_i)\ .
$$

\bigskip
\noindent
\underline{Examples:}

\begin{description}
\item{i)}
\begin{equation}
\tilde{A}_i = f(P^2) ({\bf X} \cdot {\bf P}) P_i
\end{equation}
leading to the PBs of the phase space variables
$$
\{ X_i, X_j \} = \frac{f(P^2)}{1-P^2 f(P^2)} \epsilon_{ij} L, \quad L: = \epsilon_{k\ell} X_k P_\ell
$$
\begin{equation}
\{ X_i, P_j\} = \delta_{ij} + \frac{f(P^2)}{1-P^2 f(P^2)} P_i P_j\ .
\end{equation}
A particular example is given by $f(P^2) = \frac{\Theta}{1+P^2 \Theta}$ and therefore \newline
$\frac{f}{1-P^2 f} = \Theta$.

This gives exactly Snyder's NC-algebra presented in 1947 [3].

Another case, defined by 
\begin{equation}
f(P^2) = 2/P^2,
\end{equation}
can be related to a deformed Galilei algebra (to be discussed in the next section).

\item{ii)} 
\begin{equation}
\tilde{A}_i = \tilde{A}_i ({\bf P})
\end{equation}
leading to the PBs
\begin{equation}
\{ X_i, X_j\} = \epsilon_{ij} \tilde{B} (P^2), \quad \{ X_i, P_j \} = \delta_{ij}\ .
\end{equation}
$\tilde{B}$ is the Berry curvature for the semiclassical dynamics of electrons in condensed matter (cp. [4] and the literature cited therein). 

We may generalize (15) to the most general 1st-order Lagrangian
\begin{equation}
{\cal L} = (P_i + A_i ({\bf X}, {\bf P})) \dot{X}_i + \tilde{A}_i ({\bf X}, {\bf P}) \dot{P}_i - H ({\bf  P}, {\bf X})\ .
\end{equation}
Here $A_i ({\bf X})$ describes standard electromagnetic 
interaction (cp. section 6). A particular case of a ${\bf P}$-dependent $A_i$ has been considered in [5]. A detailed discussion of this general Lagrangian, 
leading to a noncommutative structure, is still under consideration. 
\end{description}

\section{Lagrangian realization of the $\tilde{k}$-deformed Galilei algebra as a symmetry algebra}

In 1991 Jurek and his collaborators Nowicki, Ruegg and Tolstoy invented
 the $k$-deformed Poincar\'{e} algebra [6]. By rescaling the Poincar\'{e} generators 
and $k$, the corresponding nonrelativisitic limit, the $\tilde{k}$-deformed Galilei algebra, has been derived by Giller et al. [7] and, in a different basis, by Azcarraga et al. [8]. In this section we will describe a Lagrangian realization of the latter.

Again we look at the classical Lagrangian (15) specified by (18) and (20) together with the 
following choice of the Hamiltonian 
\begin{equation}
H = \tilde{k} \ln (P^2/2)\ .
\end{equation}
According to (19) we obtain the PBs 
\begin{equation}
\{ X_i, X_j \} = - \frac{2}{P^2} \epsilon_{ij} L \ \ \mbox{and} \ \ \{ X_i, P_j\} = \delta_{ij} - \frac{2}{P^2} P_i P_j
\end{equation}
which lead, together with the Hamiltonian (24), to the EOM
\begin{equation}
\dot{P}_i = 0 \ \ \ \ \mbox{and}\ \ \ \ \dot{X}_i = - \frac{2\tilde{k}}{P^2} P_i\ .
\end{equation}
Then we may define ``pseudo-boosts'' $K_i$
\begin{equation}
K_i = P_i t + \frac{P^2}{2\tilde{k}} X_i
\end{equation}
which are conserved. They satisfy, together with $P_i$ and $H$, the PB-algebra 
$$
\{ K_i, P_j\} = \frac{\delta_{ij}}{2\tilde{k}} P^2 - \frac{P_i P_j}{\tilde{k}}\ ,
$$
\begin{equation}
\{ K_i, H\} = - P_i\ , \ \ \{K_i, K_j\} = 0\ .
\end{equation}
Together with the standard algebra of translations (represented by $P_i$ and $H$)
 and rotations (represented by $L$) the relations (28) build the $\tilde{k}$-deformed Galilei algebra derived in [8].

The limit $\tilde{k} \to \infty$ leads to a divergent Hamiltonian (24). Therefore, the $\tilde{k} $-deformation does not have a standard ``no-deformation limit''. Nevertheless the PB-algebra (28) gives the standard expressions in this limit.

\section{Electromagnetic interaction and the Hall effect}

How to introduce electromagnetic (e.m.) interaction into ${\cal L}_{ext}$ (10)?

In the commutative case we have the principle of minimal e.m. coupling
\begin{equation}
P_i \dot{X}_i - \frac{P^2_i}{2m} \to (P_i + eA_i ({\bf X}, t)) \dot{X}_i 
- \frac{P_i^2}{2m} + eA_0 ({\bf X}, t),
\end{equation}
called the minimal additon rule, which is equivalent, due to the point transformation $P_i \to P_i - e A_i$, to the minimal substitution rule 
\begin{equation}
P_i \dot{X}_i - \frac{P_i^2}{2m} \to P_i \dot{X}_i - \frac{(P_i - eA_i)}{2m} + e A_0 ({\bf X}, t)\ .
\end{equation}
In the noncommutative case the equivalence of minimal addition and minimal substitution rule is not valid. 
Therefore we have to consider two different ways of introducing the minimal e.m. coupling:

\bigskip
\noindent
\underline{Minimal addition} (Duval-Horvathy [9], called \underline{DH-model}):
\begin{equation}
{\cal L}_{ext} \to {\cal L}_{e.m.} = {\cal L}_{ext} + e(A_i \dot{X}_i + A_0),
\end{equation}
which, as usual, is quasiinvariant w.r.t. standard gauge transformations
\begin{equation}
A_\mu ({\bf X}, t) \to A_\mu ({\bf X},t) + \partial_\mu \Lambda ({\bf X}, t)\ .
\end{equation}

\noindent
\underline{Minimal substitution} (Lukierski-Stichel-Zakrzewski [10], called \underline{L.S.Z.-model})\footnote{The gauge fields in this model we provide with a hat in order to distinguish them from the corresponding quantities in the DH-model.}
\begin{equation}
H_{ext} = \frac{P^2_i}{2m} \to H_{e.m.} = \frac{(P_i - e\hat{A}_i)^2}{2m} - e \hat{A}_0\ .
\end{equation}
The corresponding Lagrangian is quasiinvariant w.r.t. generalized gauge transformations, given in infin. form by 
\begin{equation}
\delta \hat{A}_\mu ({\bf X},t) : = \hat{A}_\mu^\prime ({\bf X} + \delta {\bf X}, t)- \hat{A}_\mu ({\bf X}, t) = \partial_\mu \Lambda ({\bf X}, t)
\end{equation}
with
\begin{equation}
\delta X_i = - e \Theta \epsilon_{ij} \partial_j \Lambda
\end{equation}
supplemented by
\begin{equation}
\delta P_i = e \partial_i \Lambda\ .
\end{equation}
Note, that the coordinate transformations (34) are area preserving. 

It turns out that both models are related to each other 
by a noncanonical transformation of phase space variables supplemented by
 a classical Seiberg-Witten transformation of the corresponding gauge potentials:

\noindent
If we denote the phase space variables and potentials for

-- the DH-model by $(\mbox{\boldmath$\eta, {\cal P}$}, A_\mu)$

-- the L.S.Z.-model by $({\bf X}, {\bf P}, \hat{A}_\mu)$

\noindent
then we find
\begin{equation}
\eta_i ({\bf X},t) = X_i + e \Theta \epsilon_{ij} \hat{A}_j ({\bf X}, t)
\end{equation}
\begin{equation}
{\cal P}_i = P_i - e \hat{A}_i ({\bf X},t)
\end{equation}
with the resp. field strengths related by
\begin{equation}
\hat{F}_{\mu \nu} ({\bf X},t) = \frac{F_{\mu \nu} (\mbox{\boldmath$ \eta$}, t)}{1-e\Theta B ( \mbox{\boldmath$ \eta$},t)}\ .
\end{equation}
The Seiberg-Witten transformation between the resp.~gauge fields is more involved and will not be reproduced here (for details cp.~[10]). 

\noindent
These results lead to an interesting by-product:

\noindent
Consider the PBs of coordinates in both models, given by
\begin{equation}
\{ \eta_i, \eta_j\} = \frac{\frac{\Theta}{m^2}\ \epsilon_{ij}}{1-e\Theta B( \mbox{\boldmath$ \eta$},t)} \ \ \mbox{and} \ \ \{ X_i, X_j\} = \frac{\Theta}{m^2} \epsilon_{ij}
\end{equation}
then the foregoing results implicitly give the coordinate 
transformation between a model with a constant noncommutativity parameter $\Theta$ and 
one with an arbitrary coordinate-dependent noncommutativity function $\Theta ({\bf X},t)$ (this result has been rediscovered quite recently in [11]).

\noindent
Now the question arises, which of both models has to be used for physical applications? 
Let us look at one example, the Quantum Hall effect,  in the limit of large e.m. fields. 
In the case of the DH-model [9] the Hall law
\begin{equation}
\dot{X}_i = \epsilon_{ij} \frac{E_j}{B}
\end{equation}
is valid at the critical magnetic field
\begin{equation}
B_{crit} = (e\Theta)^{-1}\ .
\end{equation}
Then it follows from the field transformation law (39) that, for the L.S.Z.-model, the Hall law is 
valid in the limit of large e.m. fields as required. In order to see this in more detail we have to 
consider the EOM for the L.S.Z.-model formulated in terms of the gauge-invariant phase 
space variables $\mbox{\boldmath$ \eta$}$ and $\mbox{\boldmath$ {\cal  P}$}$. From (29)
 and the corresponding PBs we obtain $(e=1, m=1)$
\begin{eqnarray}
\dot{\eta}_i &=& (1 + \Theta \hat{B}){\cal P}_i - \Theta \epsilon_{ij} \hat{E}_j\\
\dot{{\cal P}}_i &=& \hat{B} \epsilon_{ij} {\cal P}_j + \hat{E}_i\ .
\end{eqnarray}
For the particular case of homogeneous e.m. fields we obtain finally
\begin{equation}
\ddot{\eta}_i = \hat{B} \epsilon_{ij} \dot{\eta}_j + \hat{E}_i
\end{equation}
leading to the Hall law (41) in the high field limit. 

Note that (45) has the same functional form as in the commutative case.

\section{Supersymmetry}

In the following, we supersymmetrize the e.m.~coupling models treated in the last section.
 To do that we follow the treatment in section 3 of [12]. For that, we
consider standard $N=2$ SUSY characterized by 
\begin{equation}
H = \frac{i}{2} \{ Q, \bar{Q}\}
\end{equation}
and
\begin{equation}
\{ Q, Q\} = \{ \bar{Q}, \bar{Q}\} = 0\ .
\end{equation}
In order to construct the supercharge $Q$, satisfying (46), we start with the common structure of the bosonic Hamiltonian for both models $(e=1, m=1)$
\begin{equation}
H_b = \frac{1}{2} ({\cal P}^2_i + W_i^2 ({\bf X}))
\end{equation}
with
$$
{\cal P}_i = P_i \ \ \ \mbox{for the DH-model}
$$
and
$$
{\cal P}_i = P_i - A_i \ \ \ \mbox{for the L.S.Z.-model}. 
$$
Note that, in accordance with the quantized form of (46), the potential term in (48) is chosen to be positive
\begin{equation}
A_0 = - \frac{1}{2} W^2_i\ .
\end{equation}
In order to add to (48) its fermionic superpartner, we supplement the bosonic
 phase space variables with fermionic coordinates $\psi_i (\bar{\psi}_i)$ satisfying canonical PBs
\begin{equation}
\{ \psi_i, \bar{\psi}_j \} = - i \delta_{ij}\ .
\end{equation}
Now we assume
\begin{equation}
Q = i ({\cal P}_i + i W_i)\psi_i
\end{equation}
such that (48) is valid. But now the relations (47) are fulfilled only if the following two 
conditions are satisfied:
\begin{equation}
\{ {\cal P}_i , {\cal P}_j\} = \{ W_i, W_j\}
\end{equation}
and
\begin{equation}
\{ {\cal P}_i, W_j\} = \{ {\cal P}_j, W_i \}\ .
\end{equation}
It can be shown that (53) is satisfied automatically in both models, whereas (52) fixes the magnetic field in terms of $W_i$ (same form for both models): 
\begin{equation}
B = \frac{\Theta}{2} \epsilon_{ij} \epsilon_{k\ell} \partial_k W_i \partial_\ell W_j\ .
\end{equation}
The connection between B-field (54) and electric potential $A_0$ (49) takes a simple form in the case of rotational invariance. From
\begin{equation}
W_i ({\bf X}) = \partial_i W (r)
\end{equation}
we obtain
\begin{equation}
A_0 (r) = - \frac{1}{2} (W^\prime (r))^2
\end{equation}
and
\begin{equation}
B(r) = - \frac{\Theta}{r} A^\prime_0 (r)\ .
\end{equation}
As an example, consider the harmonic oscillator.  

\noindent
Then 
\begin{equation}
A_0 = - \frac{\omega^2}{2} r^2
\end{equation}
and we obtain a homogeneous $B$-field of strength
\begin{equation}
B = \Theta \omega^2\ .
\end{equation}

\section{Miscellaneous results}

The Galilean invariant decomposition of 6-dim phase space into invariant 
subspaces (cp. section 2) only holds for $m \neq 0$. For the case of vanishing mass 
we have to live with a 6-dim phase space as long as we keep Galilean invariance (for 
a reduction to 4-dim phase space in other (interacting) cases cp. [13]). But the $m=0$ 
model shows a higher symmetry: exotic Galilean conformal symmetry supplemented by an additional
 hidden $0(2,1)$ symmetry [14].

Other interesting results treat homogeneous e.m. fields $E_i$ and $B$ as elements of either an enlarged exotic Galilean algebrea [4] or of an enlarged Galilean conformal algebra [14]. 

It is outside the scope of this paper to discuss these interesting results in more detail.

\section{Outlook}

Dynamics on the noncommutative plane is a fascinating field. Much has
 still to be done -- hopefully in continuing my very successful collaboration with Jurek as well as with Wojtek. 

\section{Acknowledgements}

I'm grateful to J.~Azcarraga, S.~Gosh, J.~Lukierski and W.~Zakrzewski for discussions and helpful remarks.

\section{Bibliography}

\begin{description}
\item{[1]} J.~Lukierski, P.C.~Stichel and W.J.~Zakrzewski, Ann. Phys. (N.Y.) {\bf 260}, 224 (1997).

\item{[2]} P.A. Horvathy and M.S.~Plyushchay, JHEP {\bf 0206}, 033 (2002).

\item{[3]} H.S.~Snyder, Phys. Rev.~{\bf 71}, 38 (1947).

\item{[4]} P.A.~Horvathy, L.~Martina and P.C.~Stichel, Phys.~Lett.~{\bf B 615}, 87 (2005).
C.~Duval, Z.~Horvath, P.A.~Horvathy, L.~Martina and P.C.~Stichel, Mod.~Phy.~Lett.~{\bf 20}, 373 (2006); Phys.~Rev.~Lett.~{\bf 96}, 099701 (2006).

\item{[5]} S.~Gosh, Phys.~Lett.~B {\bf 638}, 350 (2006).

\item{[6]} J.~Lukierski, A.~Nowicki, H.~Ruegg and V.N.~Tolstoy, Phys.~Lett.~{\bf B 264}, 331 (1991).

\item{[7]} S.~Giller, P.~Kosinski, M.~Majewski, P.~Maslanka and J.~Kunz, Phys.~Lett.~{\bf B~286}, 57 (1992).

\item{[8]} J.A.~De Azcarraga and J.C.~P\'{e}rez Bueno, J.~Math.~Phys.~{\bf 36}, 6879 (1995).

\item{[9]} C.~Duval and P.A.~Horvathy, Phys.~Lett.~{\bf B 479}, 284 (2000); J.~Phys. {\bf A~34}, 10097 (2001).

\item{[10]} J.~Lukierski, P.C.~Stichel and W.J.~Zakrzewski, Ann.~Phys. (N.Y.) {\bf 306}, 78 (2003).

\item{[11]} C.D.~Fosco and G.~Torroba, Phys.~Rev. {\bf D 71}, 065012 (2005).

\item{[12]} J.~Lukierski, P.C.~Stichel and W.J.~Zakrzewski, Phys.~Lett.~{\bf B 602}, 249 (2004).

\item{[13]} P.C.~Stichel and W.J.~Zakrzewski, Ann.~Phys. (N.Y.), {\bf 310}, 158 (2004).

\item{[14]} J.~Lukierski, P.C.~Stichel and W.J.~Zakrzewski, Phys.~Lett. {\bf A 357}, 1 (2006).
\end{description}
\end{document}